\documentclass[aps,prb,reprint,amsmath,amssymb,superscriptaddress]{revtex4-2}
\usepackage{graphicx}
\usepackage{dcolumn}
\usepackage{bm,epsfig}
\usepackage{epstopdf}
\usepackage{xcolor}
\usepackage{float}
\usepackage{tabularx}
\usepackage{hyperref}

\usepackage[caption=false]{subfig}

\newcommand{\jan}[1]{{\color{blue}#1}}

\begin{document}

\title{Localized $f$-electron magnetism in the semimetal Ce$_{3}$Bi$_{4}$Au$_{3}$}

\author{M. O. Ajeesh} \email{ajeesh@lanl.gov} \affiliation{Los Alamos National Laboratory, Los Alamos, New Mexico 87545, USA}
\author{S. K. Kushwaha} \affiliation{National High Magnetic Field Laboratory, Los Alamos, New Mexico 87545, USA}
\author{S. M. Thomas} \affiliation{Los Alamos National Laboratory, Los Alamos, New Mexico 87545, USA}
\author{J. D. Thompson} \affiliation{Los Alamos National Laboratory, Los Alamos, New Mexico 87545, USA}
\author{M. K. Chan} \affiliation{National High Magnetic Field Laboratory, Los Alamos, New Mexico 87545, USA}
\author{N. Harrison} \affiliation{National High Magnetic Field Laboratory, Los Alamos, New Mexico 87545, USA}
\author{J. M. Tomczak} \affiliation{Department of Physics, King’s College London, Strand, London WC2R 2LS, United Kingdom} \affiliation{Institute of Solid State Physics, TU Wien, Vienna 1040, Austria}
\author{P. F. S. Rosa} \affiliation{Los Alamos National Laboratory, Los Alamos, New Mexico 87545, USA}

\date{\today}

\begin{abstract}

Ce$_{3}$Bi$_{4}$Au$_{3}$ crystallizes in the same non-centrosymmetric cubic structure as the prototypical Kondo insulator Ce$_{3}$Bi$_{4}$Pt$_{3}$.
Here we report the physical properties of Ce$_{3}$Bi$_{4}$Au$_{3}$ single crystals using magnetization, thermodynamic, and electrical-transport measurements. Magnetic-susceptibility and heat-capacity data reveal antiferromagnetic (AFM) order below $T_N=3.2$~K. The magnetic entropy $S_{\rm mag}$ reaches $R$ln2 slightly above $T_N$, which suggests localized $4f$-moments in a doublet ground state. Multiple field-induced magnetic transitions are observed at temperatures below $T_N$, which indicate a complex spin structure with competing interactions. Ce$_{3}$Bi$_{4}$Au$_{3}$ shows semimetallic behavior in electrical resistivity measurements in contrast to the majority of reported Cerium-based 343 compounds. Electrical-resistivity measurements under hydrostatic pressure reveal a slight enhancement of $T_N$ under pressures up to 2.3~GPa, which supports a scenario wherein Ce$_{3}$Bi$_{4}$Au$_{3}$ belongs to the far left of the Doniach phase diagram dominated by Ruderman–Kittel–Kasuya–Yosida (RKKY) interactions. Using realistic many-body simulations, we confirm the semi-metallic electronic structure of Ce$_{3}$Bi$_{4}$Au$_{3}$ and quantitatively reproduce its local moment behavior in the paramagnetic state.

\end{abstract}

\maketitle

\section{Introduction}

Cerium-based ternary compounds crystallizing in the cubic Ce$_3X_4T_3$ ($X=$ Bi, Sb; $T =$ transition metal) structure present a wide variety of ground-state properties owing to the delicate interplay between the crystal lattice, the electron filling, and the hybridization between $f$-electrons and the conduction electron sea. 
For instance, the isoelectronic compounds Ce$_3$Bi$_4$Pt$_3$ and Ce$_3$Sb$_4$Pt$_3$ exhibit Kondo-insulating behavior in an intermediate valence regime~\cite{Hundley90, Fisk95, Peter00, NGCS}, whereas the ground state of Ce$_{3}$Bi$_{4}$Pd$_{3}$ has been argued to be either a Weyl-Kondo semimetal or a Kondo insulator with narrower gap compared to Ce$_3$Bi$_4$Pt$_3$~\cite{Dzsaber17, Lai18, Dzsaber21, Kushwaha19, Ajeesh21,jmt_radialKI}.

In contrast, substitution by group 11 transition metals, such as Cu or Au, tends to drive Ce ions into a localized $3+$ state and tune the material towards a band insulating/semimetallic state~\cite{Kenichi96, Jones99, Watcher99}. Though Ce$_{3}$Sb$_{4}$Au$_{3}$ presents semiconducting behavior seemingly similar to that of the Pt counterpart, the origin of the gap cannot be Kondo hybridization. Instead, electron count through the Zintl concept [3Ce$^{3+}+ $4Sb$^{3-}+ $3Au$^{1+} = +9 -12 +3$] is suggestive of a conventional semiconducting gap~\cite{Seibel16}. 
Interestingly, Ce$_{3}$Sb$_{4}$Au$_{3}$ shows a large specific heat coefficient at low temperatures, which is unexpected in a semiconductor with localized $f$ electrons. This unusual behavior has been attributed to the presence of an $f$-level resonant state within a conventional band gap~\cite{Lee07, Lee08}. In the case of Ce$_{3}$Sb$_{4}$Cu$_{3}$, an activated behavior in the electrical resistivity with decreasing temperature was initially taken as evidence for a semiconducting gap~\cite{Patil96}. Later studies using optical reflectivity and Hall measurements, however, argued against the presence of an energy gap, and the temperature dependence of the electrical resistivity was ascribed to changes in the carrier mobility, categorizing Ce$_{3}$Sb$_{4}$Cu$_{3}$ as a semimetal~\cite{Watcher99}. Finally, Ce$_{3}$Sb$_{4}$Au$_{3}$ does not order magnetically, whereas Ce$_{3}$Sb$_{4}$Cu$_{3}$ undergoes magnetic ordering at 10~K to a canted antiferromagnetic (AFM) phase~\cite{Watcher00, Herrmann99, Schnelle01}.

Ce$_{3}$Bi$_{4}$Au$_{3}$, a less studied compound in the 343 family, also crystallizes in the cubic Y$_{3}$Sb$_{4}$Au$_{3}$-type structure (space group $I\overline{4}3d$)~\cite{Seibel16}. Magnetic susceptibility measurements reveal a Curie-Weiss behavior consistent with Ce$^{3+}$ local moments which undergo antiferromagnetic ordering at $T_N=2.9$~K~\cite{Seibel16}. Electrical-resistivity measurements in $RE_{3}$Bi$_{4}$Au$_{3}$ ($RE=$ La, Nd, Sm) revealed a weak decrease in resistivity on cooling with a corresponding residual resistivity ratio ($\rho_{\mathrm{300~K}}/\rho_{\mathrm{10~K}}$) of only about 1.3--1.4~\cite{Seibel16}. This poor metallic behavior was taken as indication that members of the $RE_{3}$Bi$_{4}$Au$_{3}$ family are doped semiconductors with largely temperature-independent electrical resistivities. Notably, these materials show a large Seebeck value, which holds promise for thermoelectric applications~\cite{Witas21,Seibel16, Young99}.

In this article, we investigate the physical properties of Ce$_{3}$Bi$_{4}$Au$_{3}$ single crystals by means of magnetic, thermodynamic, and electrical-transport measurements. The magnetic ground state of Ce$_{3}$Bi$_{4}$Au$_{3}$ is characterized in detail along with the effect of magnetic field and hydrostatic pressure. Our results reveal localized $f$ moments in a crystalline electric field doublet ground state. Under hydrostatic pressure, $T_N$ slowly increases, which indicates Ce$_{3}$Bi$_{4}$Au$_{3}$ is dominated by RKKY interactions that lead to its AFM ground state. Many-body calculations, performed in the framework of density functional theory plus dynamical mean field theory (DFT+DMFT), reveal the presence of a small electron pocket at the Fermi level in the paramagnetic phase, which confirms its semimetallic nature. The simulations further yield a Curie-Weiss magnetic susceptibility in quantitative agreement with experiments.
Together, our combined experimental and theoretical results enable a broader comparison between Ce-343 compounds and shed light on the distinct ground states in this family.

\section{Methods}

Single crystals of Ce$_{3}$Bi$_{4}$Au$_{3}$ and La$_{3}$Bi$_{4}$Au$_{3}$ were grown by the Bi-flux method with a starting composition Ce(La):Au:Bi= 1:2:15. The starting materials were loaded into an alumina crucible and sealed in an evacuated quartz tube. The quartz tube was then heated to $850^{\circ}$C at $85^{\circ}$C/h and held at this temperature for 36 hours, followed by a multi-step cool down; to $750^{\circ}$C in 1 hour, then to $550^{\circ}$C in 100 hours, and finally to $400^{\circ}$C in 50 hours. Excess Bi flux was removed by centrifugation after heating the quartz tube to $450^{\circ}$C. Single crystals of polyhedral shape with largest dimension of about 1.5~mm were obtained. The crystallographic structure was verified at room temperature by a Bruker D8 Venture single-crystal diffractometer equipped with Mo radiation.

Magnetization measurements were carried out in the temperature range $1.8~{\rm K}-350$~K and in magnetic fields to 7~T using a SQUID-VSM magnetometer (MPMS3, Quantum Design). High-field magnetization measurements to 30~T in pulsed magnetic fields were performed at the National High Magnetic Field Laboratory, Los Alamos, USA. The heat capacity was measured in a Quantum Design Physical Property Measurement System (PPMS) using a calorimeter that utilizes a quasi-adiabatic thermal relaxation technique. Electrical-resistivity measurements were performed on polished crystals in a standard 4-terminal method wherein electrical contacts to the samples were made using 12.5 $\mu$m platinum wires and silver paste. Electrical-transport measurements under hydrostatic pressure were carried out using a double-layered piston-cylinder-type pressure cell with Daphne 7373 oil as the pressure-transmitting medium. The pressure inside the sample space was determined at low temperatures by the shift of the superconducting transition temperature of a piece of lead (Pb). Electrical resistivity was measured using an AC resistance bridge (Model 372, Lake Shore) at a measuring frequency of 13.7 Hz together with a PPMS. 

For the many-body DFT+DMFT simulations, we initialized the crystal structure of Ce$_3$Bi$_4$Au$_3$ with atomic positions from Ref.~\cite{Seibel16} and the experimental lattice constant, $a=10.221$\AA, of the current work. Within the confines of space-group $I\bar{4}3d$, only the position ($u$,$u$,$u$) of the Bi site is not pre-determined.
We relax this coordinate within density-functional theory (DFT; with the PBE functional) using wien2k~\cite{wien2k2020}. DMFT calculations were then performed in the same realistic setting (DFT+DMFT)
as previously for Ce$_3$Bi$_4$Pt$_3$~\cite{LRT_Tstar,jmt_CBP_arxiv} with the code of Haule {\it et al.}~\cite{PhysRevB.81.195107},
which includes spin-orbit coupling, charge self-consistency, and full Coulomb interactions parameterized by a Hubbard $U=5.5$~eV and a Hund's $J=0.68$~eV. The impurity problem was solved using continuous-time quantum Monte Carlo in the hybridization expansion and the Ce-$4f$ hybridization was accounted for in an energy window of $\pm 8$~eV around the Fermi level.
Results shown use the nominal double-counting scheme~\cite{PhysRevB.81.195107}.
Magnetic degrees of freedom are assessed by sampling the local (impurity) spin susceptibility.
Analytical continuation was performed as described in Ref.~\cite{PhysRevB.81.195107}.
Transport properties were simulated in linear-response, following the methodology of Ref.~\cite{jmt_fesi}.

\section{Results and Discussion}

\subsection*{Magnetic susceptibility and magnetization}\label{susceptibility}

\begin{figure}[tb]
\centering
\includegraphics[width=1\linewidth]{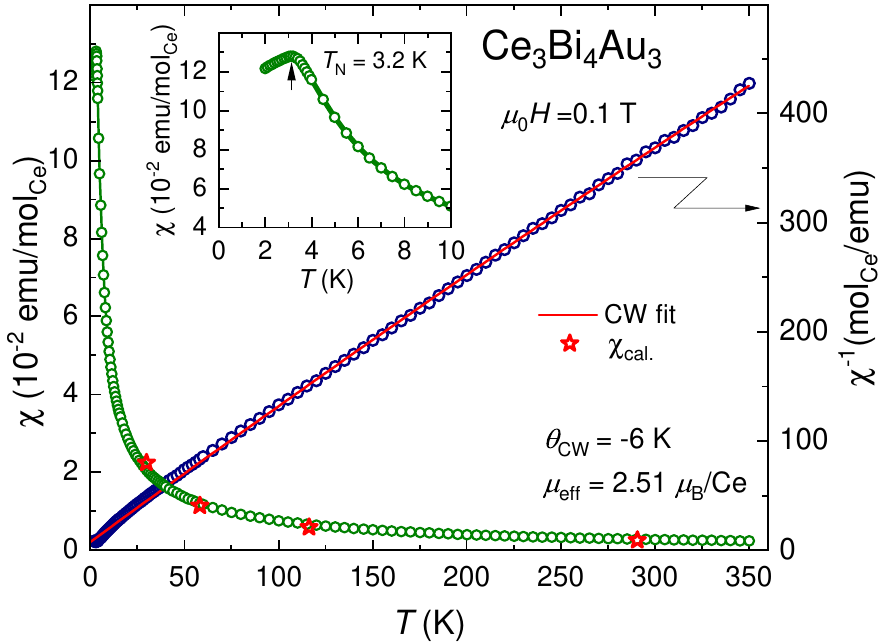}
\caption{Temperature dependence of the magnetic susceptibility $\chi=M/H$ (left axis) of Ce$_{3}$Bi$_{4}$Au$_{3}$ measured under magnetic field  $\mu_0H=0.1$~T. The inverse magnetic susceptibility $\chi^{-1}(T)$ is plotted on the right axis. The solid red line is a fit using Curie-Weiss law in the temperature interval $100{\rm~K}\leq T\leq 350{\rm~K}$. The inset presents zoom in of the low-temperature $\chi(T)$ highlighting the peak at 3.2~K corresponding to the antiferromagnetic transition.
The red star symbol stands for the local susceptibility calculated from electronic structure simulations.}
\label{ChivsT}
\end{figure}

The temperature-dependent magnetic susceptibility, $\chi(T)$, of Ce$_{3}$Bi$_{4}$Au$_{3}$ measured in an applied magnetic field of $\mu_0H=0.1$~T is depicted in Fig.~\ref{ChivsT}. At high temperatures, Ce$_{3}$Bi$_{4}$Au$_{3}$ shows paramagnetic behavior, and its susceptibility follows the Curie-Weiss (CW) law. A CW fit to the $\chi^{-1}(T)$ data (Fig.~\ref{ChivsT}, right axis) using $\chi(T) = \chi_0+C/(T-\theta_{\rm W})$ in the temperature range $100{\rm~K}\leq T\leq 350{\rm~K}$ yields $\chi_0=1.3(2)\times10^{-4}$~emu/mol$_{\rm Ce}$, the Curie constant $C=0.79(1)$~emu K/mol$_{\rm Ce}$, and the Weiss temperature $\theta_{\rm W}=-6(1)$~K. The effective moment, $\mu_{\rm eff}=\sqrt{8C}=2.51~\mu_{B}$, is close to the expected value of $2.54~\mu_{B}$ for a free Ce$^{3+}$ ion. 
The negative value of $\theta_{\rm W}$ indicates predominantly antiferromagnetic interaction between the moments. The deviation of $\chi(T)$ from CW behavior for temperatures below $\approx75$~K is likely related to crystalline electric field effects as discussed below. At low temperatures, Ce$_{3}$Bi$_{4}$Au$_{3}$ orders antiferromagnetically at around $T_{\rm N}=3.2$~K as evidenced by a cusp-like feature in $\chi(T)$ (see inset of Fig.~\ref{ChivsT}). The magnetic transition temperature as well as the CW parameters are in good agreement with previously reported values~\cite{Seibel16}.

\begin{figure}[tb]
\centering
\includegraphics[width=1\linewidth]{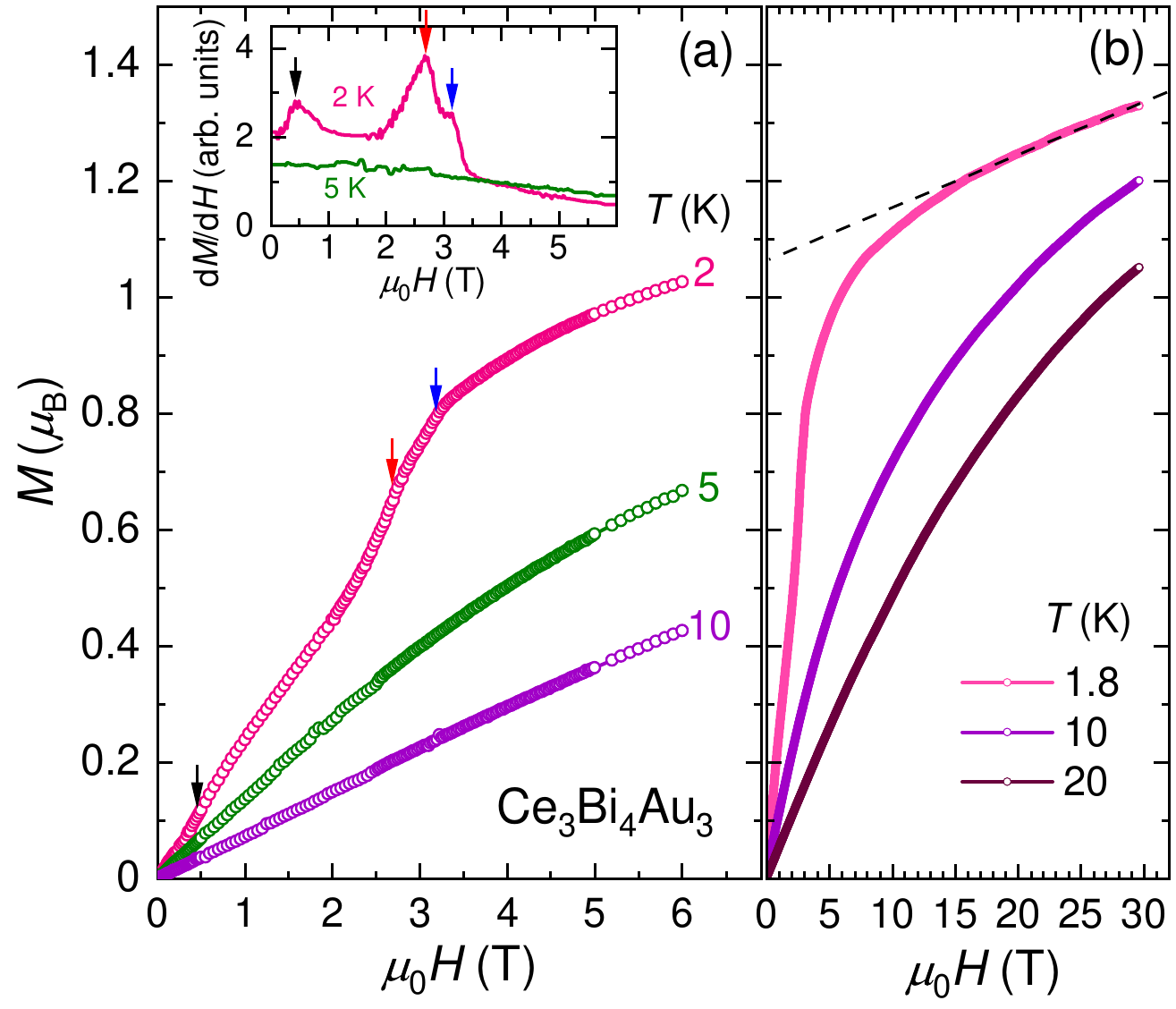}
\caption{(a) Magnetization  of Ce$_{3}$Bi$_{4}$Au$_{3}$ versus applied magnetic field, measured at different temperatures. The derivative d$M$/d$H$ is plotted in the inset. Various features in $M(H)$ are indicated by the arrows. (b) High-field magnetization measured at several temperatures.}
\label{MvsH}
\end{figure}

The magnetization of Ce$_{3}$Bi$_{4}$Au$_{3}$ as a function of applied field, $M(H)$, measured at different temperatures below and above $T_N$ is plotted in Fig.~\ref{MvsH}. At $T=2$~K, $M(H)$ displays several slope changes, which are likely associated with changes in the magnetic structure. Three noticeable features at $\mu_0H=0.5$, 2.6, and 3.2~T (marked by arrows) are detected in derivative d$M$/d$H$ curves plotted in the inset of Fig.~\ref{MvsH}a. These multi-step spin-reorientation transitions have been observed in other Ce-based materials and are a telltale sign of complex spin structures arising from competing exchange interactions ($i.e.$, magnetic frustration)~\cite{Boehm79,Thomas16}. Figure~\ref{MvsH}b shows the isothermal magnetization for magnetic fields to 30~T. Here, the magnetization data is scaled to match the low-field data from MPMS measurements. At $T=1.8$~K, $M$ reaches a value of 1.3~$\mu_{B}$/Ce at $\mu_0H=30$~T which is much smaller then the full saturation moment of the $J=5/2$ multiplet. Similar magnetization values are also observed for isostructural compounds Ce$_{3}$Sb$_{4}$Au$_{3}$~\cite{Kasaya94, Adroja95, Lee07, Beak10} and Ce$_{3}$Sb$_{4}$Cu$_{3}$~\cite{Watcher99, Schnelle01} whose crystalline electric field (CEF) ground states are Kramers doublets.

\subsection*{Heat capacity}\label{heatcapacity}
\begin{figure}[tb]
\centering
\includegraphics[width=1\linewidth]{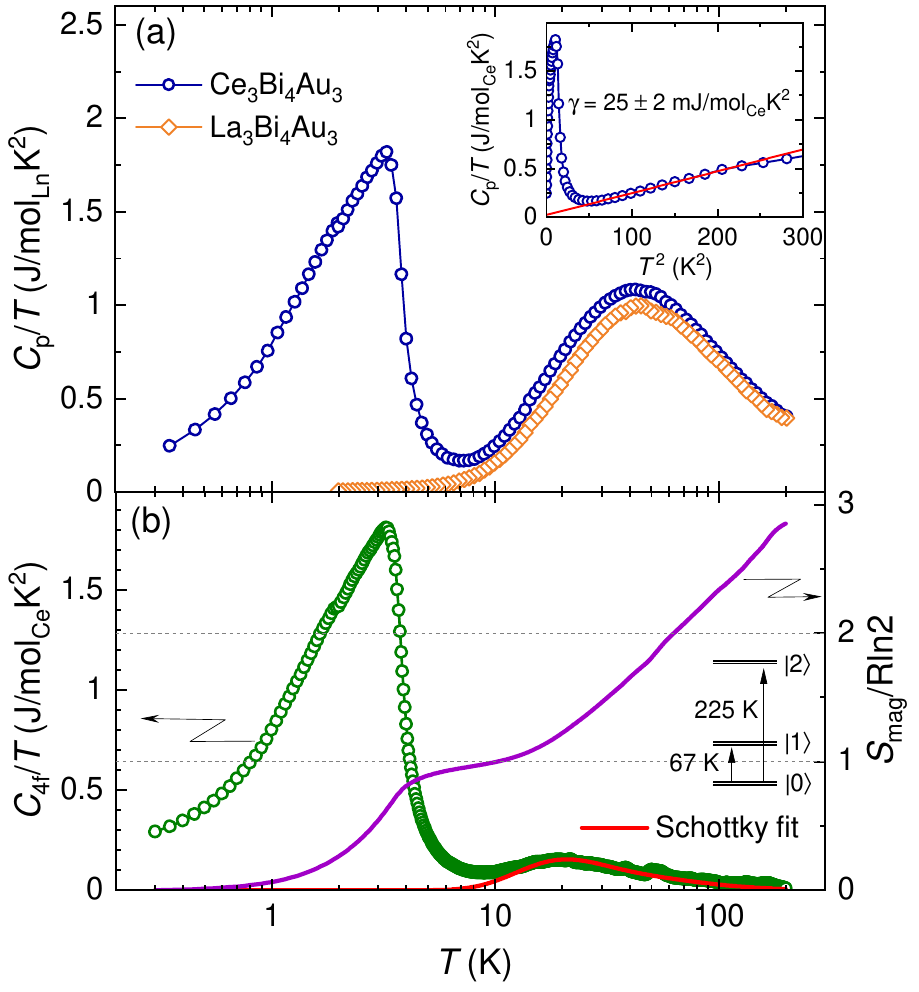}
\caption{(a) Temperature dependence of the heat capacity of Ce$_{3}$Bi$_{4}$Au$_{3}$ and the non-magnetic iso-structural compound La$_{3}$Bi$_{4}$Au$_{3}$, plotted as $C_p/T$ versus log$T$. Inset shows $C/T$ versus $T^2$ plot for Ce$_{3}$Bi$_{4}$Au$_{3}$. The solid red line is a linear fit to the data in the temperature range $8-14$~K. (b) The magnetic contribution to the heat capacity, plotted as $C_{4f}(T)/T$ versus $T$ (left axis). $C_{4f}$ is estimated by subtracting the heat capacity of the non-magnetic reference compound La$_{3}$Bi$_{4}$Au$_{3}$ from that of Ce$_{3}$Bi$_{4}$Au$_{3}$. The solid red line is a Schottky fit to the data for $10{\rm~K}\leq T\leq 200{\rm~K}$ assuming three doublet states. The calculated magnetic entropy $S_{\rm mag}(T)$ is displayed in the unit of $R\ln2$ (right axis).}
\label{CpvsT}
\end{figure}
\begin{figure}[t]
\centering
\includegraphics[width=1\linewidth]{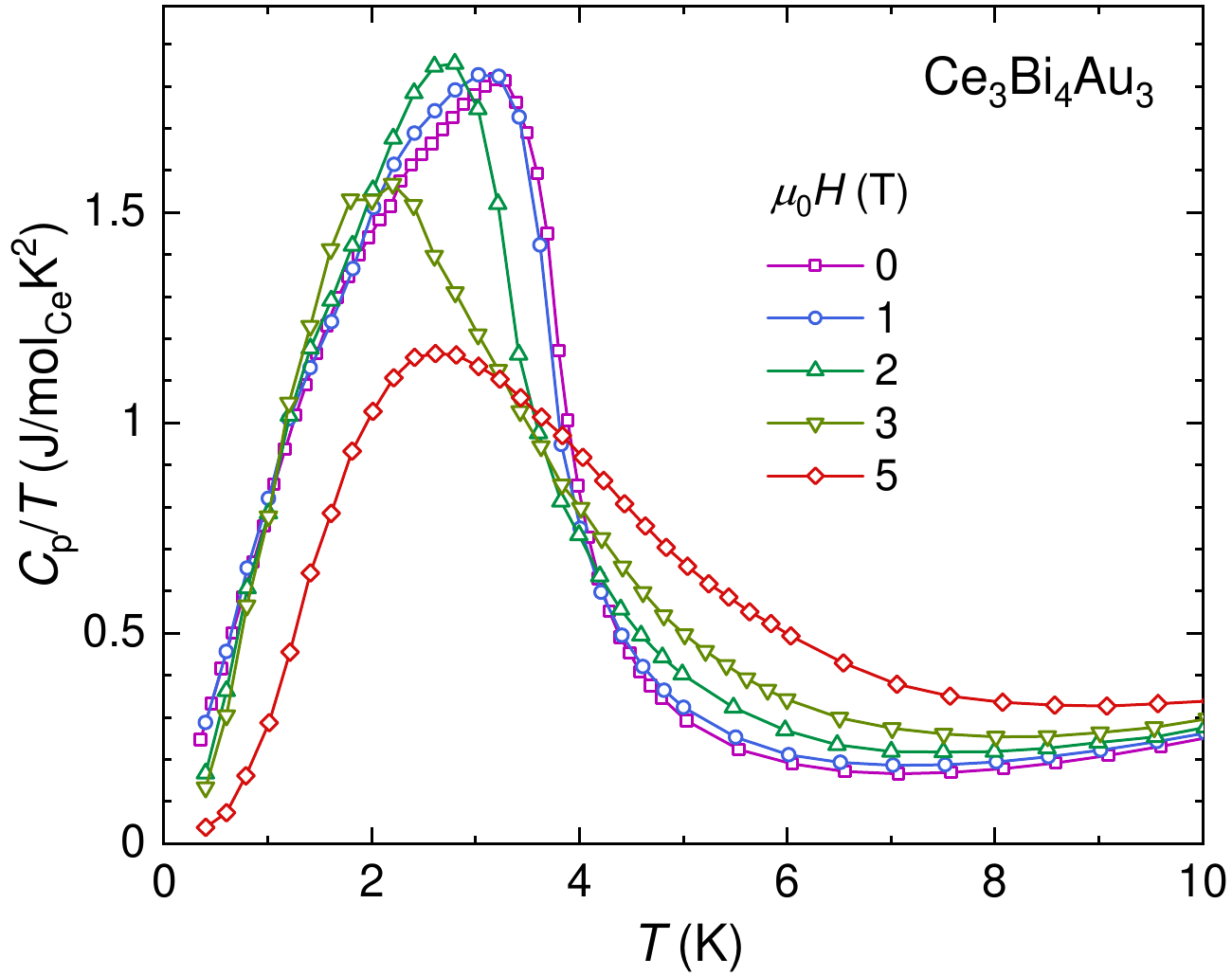}
\caption{$C_p/T$ vs. $T$ of Ce$_{3}$Bi$_{4}$Au$_{3}$ measured under different applied magnetic fields.}
\label{CpvsT-H}
\end{figure}
Figure~\ref{CpvsT}a presents the temperature dependence of the heat capacity of Ce$_{3}$Bi$_{4}$Au$_{3}$ and, as a reference, the non-magnetic iso-structural compound La$_{3}$Bi$_{4}$Au$_{3}$, plotted as $C_p/T$ versus log$T$. The sharp peak in $C_p/T$ of Ce$_{3}$Bi$_{4}$Au$_{3}$ at $T = 3.2$~K corresponds to the AFM ordering and is consistent with the magnetic-susceptibility data. In the inset of Figure~\ref{CpvsT}a, the heat capacity of Ce$_{3}$Bi$_{4}$Au$_{3}$ is plotted as $C_p/T$ versus $T^2$. There, the linear region observed between $T=8$~K and 14 K can be fit to $C_p(T) = \gamma T + \beta T^3$. The extracted Sommerfeld coefficient $\gamma=25\pm2$~mJ/mol$_{\rm Ce}$K$^2$ implies rather weak electron-correlation effects. 

The magnetic contribution to the heat capacity of Ce$_{3}$Bi$_{4}$Au$_{3}$ is estimated by subtracting the heat capacity of the non-magnetic reference compound La$_{3}$Bi$_{4}$Au$_{3}$ (see Fig.~\ref{CpvsT}b). The magnetic entropy $S_{\rm mag}$, obtained by integrating $C_{4f}/T$ over $T$, reaches about 80\% of $R$ln2 at $T_N$ and presents a small plateau before continuously increasing with further increase in temperature. This behavior corroborates a CEF doublet ground state which is well separated from the excited CEF levels. The large value of $S_{\rm mag}$ near $T_N$ points to the localized nature of the Ce moments and a rather weak Kondo effects in Ce$_{3}$Bi$_{4}$Au$_{3}$. We note that the full magnetic entropy of $R$ln2 for a doublet state is recovered only at around 10~K, which can be ascribed to the formation of short-range correlations leading up to the AFM long-range order. Within the mean-field approximation, the jump in heat capacity $\Delta C_{4f}$ at $T_N$ can be used to estimate the Kondo scale~\cite{Bredl78, Blanco94}. Here, $\Delta C_{4f}=5.86$J/mol$_{\rm Ce}$K at $T_N$, yielding a Kondo temperature of $T_K\approx4$~K. We note that this is probably an overestimation of $T_K$ due to the broad nature of the heat capacity anomaly; however, this estimate also reflects the weak Kondo effect in Ce$_{3}$Bi$_{4}$Au$_{3}$.  

Further insights into the CEF scheme of Ce$_{3}$Bi$_{4}$Au$_{3}$ can be obtained from $C_{4f}(T)$ data in the paramagnetic (PM) region, which reveals a broad hump centered around 20~K reminiscent of a Schottky anomaly associated with the excited CEF levels. Even though the overall crystal structure of Ce$_{3}$Bi$_{4}$Au$_{3}$ has cubic symmetry, the Ce sites have tetragonal point symmetry due to the  distorted tetrahedral arrangement of surrounding Au atoms. In the presence of a CEF with a tetragonal symmetry, the $6-$fold degenerate levels of Ce$^{3+}$ ion with total angular momentum $J=5/2$ split into three Kramers doublets which are energetically separated from each other. 
Therefore the high-temperature part of the $C_{4f}(T)$ data can be fitted with a Schottky contribution for 3-level system as
\begin{equation}\label{Schottky}
  C_{\mathrm Sch}(T) = \frac{R}{T^2}\frac{\sum_{i}\sum_{j}g_ig_j\Delta_i(\Delta_i-\Delta_j)e^{\frac{-(\Delta_i+\Delta_j)}{T}}}{(\sum_{i}g_ie^{\frac{-\Delta_i}{T}})^2},
\end{equation}
where $R$ is the gas constant, $i,j=0,1,2$, $g_i$ is the degeneracy of each level, and $\Delta_i$ is the energy separation from the ground state. Experimental data for Ce$_{3}$Bi$_{4}$Au$_{3}$ in the temperature range $10-200$~K are well reproduced by Eq.~\ref{Schottky} with the first and second excited levels at $67\pm0.3$ and $225\pm2$~K, respectively. A schematic representation of the CEF levels is depicted in the inset of Fig.~\ref{CpvsT}b. Note that a similar CEF scheme has also been reported for the isostructural compounds Ce$_{3}$Sb$_{4}$Au$_{3}$~\cite{Kasaya94, Adroja95, Lee07, Beak10} and Ce$_{3}$Sb$_{4}$Cu$_{3}$~\cite{Watcher99, Schnelle01}.

$C_p/T$ of Ce$_{3}$Bi$_{4}$Au$_{3}$ as a function of temperature for several applied magnetic fields is shown in Fig.~\ref{CpvsT-H}. Initially, the peak in $C_p/T$ shifts to lower temperatures with increasing field, a behaviour typical of AFM ordering. For fields above 3~T, the peak in $C_p/T$ becomes much broader and shifts to higher temperatures with further increase in field. This is consistent with the field-induced spin reorientation observed in the magnetization data. The broad hump in $C_p/T$ under higher fields corresponds to the crossover from the paramagnetic to the field-polarized phase. 
\subsection*{Electrical transport}
\label{resistivity}

\begin{figure}[tb] 
\centering
\includegraphics[width=1\linewidth]{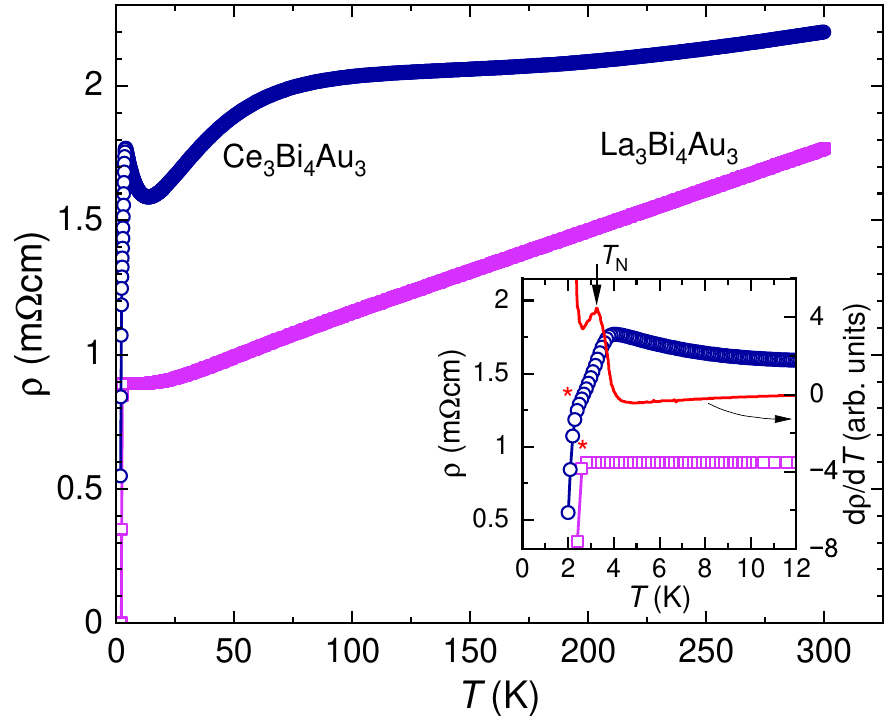}
\caption{(a) Temperature dependence of the electrical resistivity $\rho(T)$ of Ce$_{3}$Bi$_{4}$Au$_{3}$ and La$_{3}$Bi$_{4}$Au$_{3}$. The inset shows an enlarged view of the low-temperature part of $\rho(T)$. The temperature derivative of resistivity d$\rho$/d$T$ of Ce$_{3}$Bi$_{4}$Au$_{3}$ is plotted in the right axis. Arrow indicates the anomaly corresponding to AFM ordering in Ce$_{3}$Bi$_{4}$Au$_{3}$ and the star symbols denote the superconducting transitions possibly coming from Au$_2$Bi inclusions.}
\label{rhovsT}
\end{figure}

The electrical resistivity, $\rho(T)$, of Ce$_{3}$Bi$_{4}$Au$_{3}$ and La$_{3}$Bi$_{4}$Au$_{3}$ are shown in Fig.~\ref{rhovsT}. Both materials exhibit semimetallic behavior with room temperature resistivities in the range of 1-3 m$\Omega$cm. $\rho(T)$ appear to be sample dependent, probably due to the presence of inclusions of unwanted phases due to excess flux used in the sample growth. For instance, the resistivity of polycrystalline La$_{3}$Bi$_{4}$Au$_{3}$ reported earlier is an order of magnitude larger than that we obtained here~\cite{Seibel16}. The contributions from impurity inclusions in our crystals could not be removed even after carefully polishing the samples, which implies that the inclusions may be present within the bulk of the crystals. Effects of impurity inclusions on the magnetotransport data is discussed in Appendix A. Similar issues of impurity inclusions in electrical-transport data have been reported in other members of the 343-compounds~\cite{Lee07, Ajeesh21}.   

The resistivity of Ce$_{3}$Bi$_{4}$Au$_{3}$ presents a broad curvature centered around $\approx70$~K, which is arguably attributed to additional scattering originating from thermal population of excited CEF levels. Note that the temperature scale corresponding to the broad feature coincides with the CEF splitting obtained from heat-capacity data. At low temperatures, another increase in $\rho(T)$ is observed below 10~K followed by a pronounced kink at about 3.2~K due to the loss of spin-disorder scattering below $T_N$. The upturn in resistivity below 10~K coincides with the recovery of $R$ln2 entropy in heat capacity and likely corresponds to the onset of short-range interactions and/or Kondo scattering. Here, $T_N$ is determined from the maximum in the temperature derivative of resistivity d$\rho$/d$T$ (marked by the arrow in the inset of Fig.~\ref{rhovsT}). $\rho(T)$ of both Ce$_{3}$Bi$_{4}$Au$_{3}$ and La$_{3}$Bi$_{4}$Au$_{3}$ show a sudden drop below $T\approx2.2$~K, which is likely due to the inclusion of superconducting Au-Bi binaries, such as Au$_2$Bi~\cite{Roberts76}. The lack of any associated feature in either magnetic susceptibility or heat capacity suggests that the impurity volume fraction is very small.

\subsection*{Effect of hydrostatic pressure}

\begin{figure}[tb]
  \centering
  \includegraphics[width=1\linewidth]{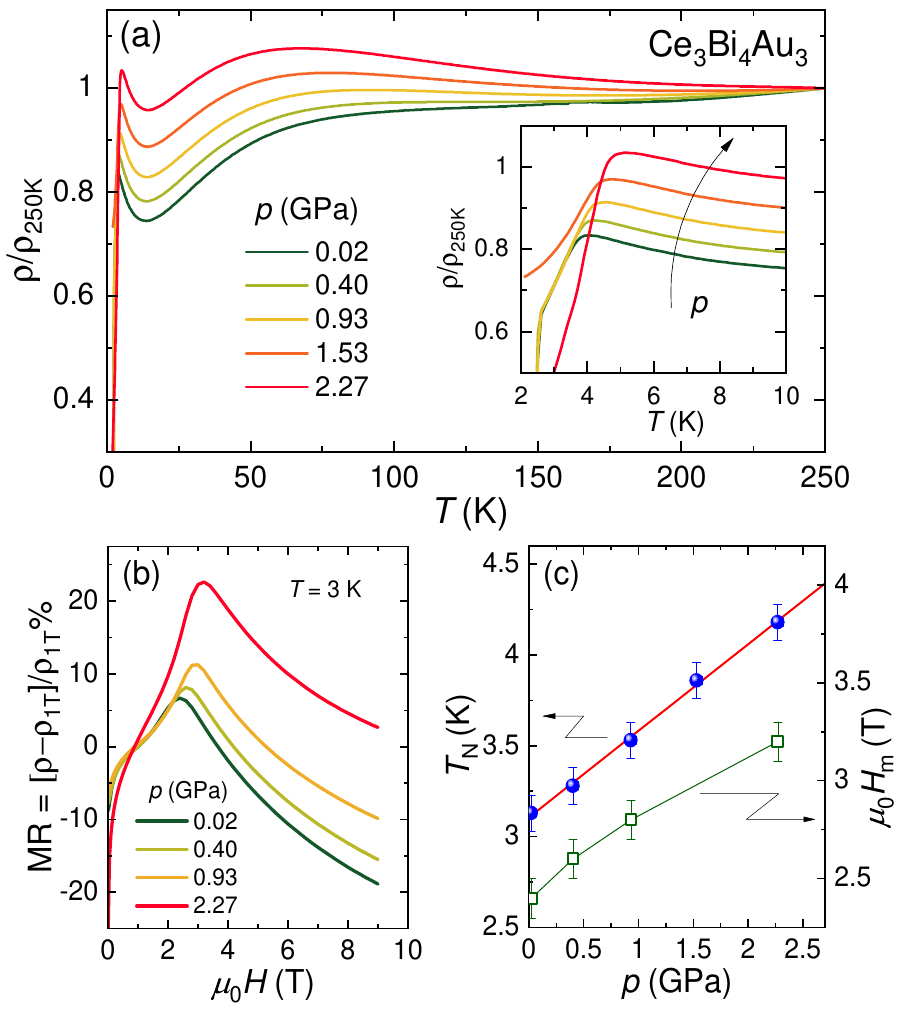}
  \caption{(a) Normalized resistivity $\rho/\rho_{\rm 250 K}$ of Ce$_{3}$Bi$_{4}$Au$_{3}$ vs. $T$ under several applied pressures. The inset shows an enlarged view of the low-temperature region of the curves. (b) Magnetoresistance MR$=[\rho-\rho_{\rm 1 T}]/\rho_{\rm 1 T}$ as a function of magnetic field measured at $T=3$~K and different pressures. (c) Pressure dependence of the AFM transition temperature $T_N$ (left axis) and the spin-reorientation critical field $H_m$ (right axis). The red line is a linear fit to $T_N(p)$ data.}
  \label{RvsT-P}
  \end{figure}

The evolution of the magnetic transition in Ce$_{3}$Bi$_{4}$Au$_{3}$ is studied using electrical-transport measurements under hydrostatic pressure. The temperature dependence of the normalized electrical resistivity $\rho/\rho_{\rm 250 K}$ for several applied pressures are plotted in Fig.~\ref{RvsT-P}a. The anomaly in resistivity corresponding to the AFM transition slightly shifts to higher temperatures with increasing pressure and reaches 4.6~K at $p=2.27$~GPa (see inset of Fig.~\ref{RvsT-P}a). $T_N$, estimated from the maximum in the corresponding d$\rho$/d$T$ curve, as a function of pressure is plotted in Fig.~\ref{RvsT-P}c. $T_N$ increases linearly with pressure and a linear fit to the data yields the slope d$T_N$/d$p=0.48\pm0.01$~K/GPa. This enhancement in $T_N$ is consistent with the Doniach phase diagram wherein the behavior of $T_N(p)$ is described by the competition between Ruderman–Kittel–Kasuya–Yosida (RKKY) and Kondo interactions~\cite{Doniach77}. The strongly localized nature of the $4f$ moments and weak Kondo scale place Ce$_{3}$Bi$_{4}$Au$_{3}$ in the far left of the Doniach phase diagram dominated by RKKY interaction. In this region, the increase in hybridization strength with pressure predominantly enhances the RKKY interaction and hence the magnetic order. Therefore, much higher pressure may be needed to suppress magnetism in Ce$_{3}$Bi$_{4}$Au$_{3}$.   

The relative change in transverse magnetoresistance MR$=[\rho-\rho_{\rm 1 T}]/\rho_{\rm 1 T}$ as a function of magnetic field measured at $T=3$~K for several pressures is displayed in Fig.~\ref{RvsT-P}b. Here, MR($H$) is obtained by symmetrizing the resistivity data measured in positive and negative magnetic fields to correct for the misalignment of electrical contacts. MR($H$) shows a steep increase for fields up to 1~T. This is possibly due to the presence of excess Bi or Au-Bi binaries in the samples. The fact that this feature persists up to temperatures much higher than $T_N$ and also present in non-magnetic La$_{3}$Bi$_{4}$Au$_{3}$ (see Fig.~\ref{MRvsH} in the Appendix) supports an extrinsic origin. A detailed discussion on the effect of impurity inclusions on the magnetotransport is provided in Appendix A. It is also worth noting that very similar positive MR for fields up to 1~T is reported in flux grown samples of (Ce, La)$_{3}$Bi$_{4}$Pt$_{3}$~\cite{Hundley93}. A common feature in all these materials is the possibility of Bi-related inclusions in the samples. Nevertheless, the MR data above 1~T reflect the intrinsic properties of Ce$_{3}$Bi$_{4}$Au$_{3}$, namely, positive MR followed by a sudden decrease at the spin-reorientation transition. The initial positive MR can be explained by the cyclotron motion of the conduction electrons~\cite{Yamada73}. The sudden decrease in MR at the spin-reorientation transition is due to the loss of spin-disorder scattering as the system goes from AFM to the field-polarized phase. The critical magnetic field for this transition monotonously increases with increasing pressure, as shown in the right axis of Fig.~\ref{RvsT-P}c. This further substantiates that the AFM phase is stabilized by the application of pressures up to 2.3~GPa. Another noticeable feature in the pressure evolution of the $\rho(T)/\rho_{\rm 250 K}$ curves is that the broad hump in resistivity associated with the scattering from the excited CEF level gradually shifts to lower temperatures with increasing pressure. This would imply that the CEF scheme is modified by the application of pressure and the energy gap between the CEF ground state and the excited levels is slightly lowered by increasing pressure.

\subsection*{Electronic structure simulations}

\begin{figure*}[tb]
\hspace{-0.75cm}
  \begin{minipage}{0.5\textwidth}
{\includegraphics[width=1.25\linewidth]{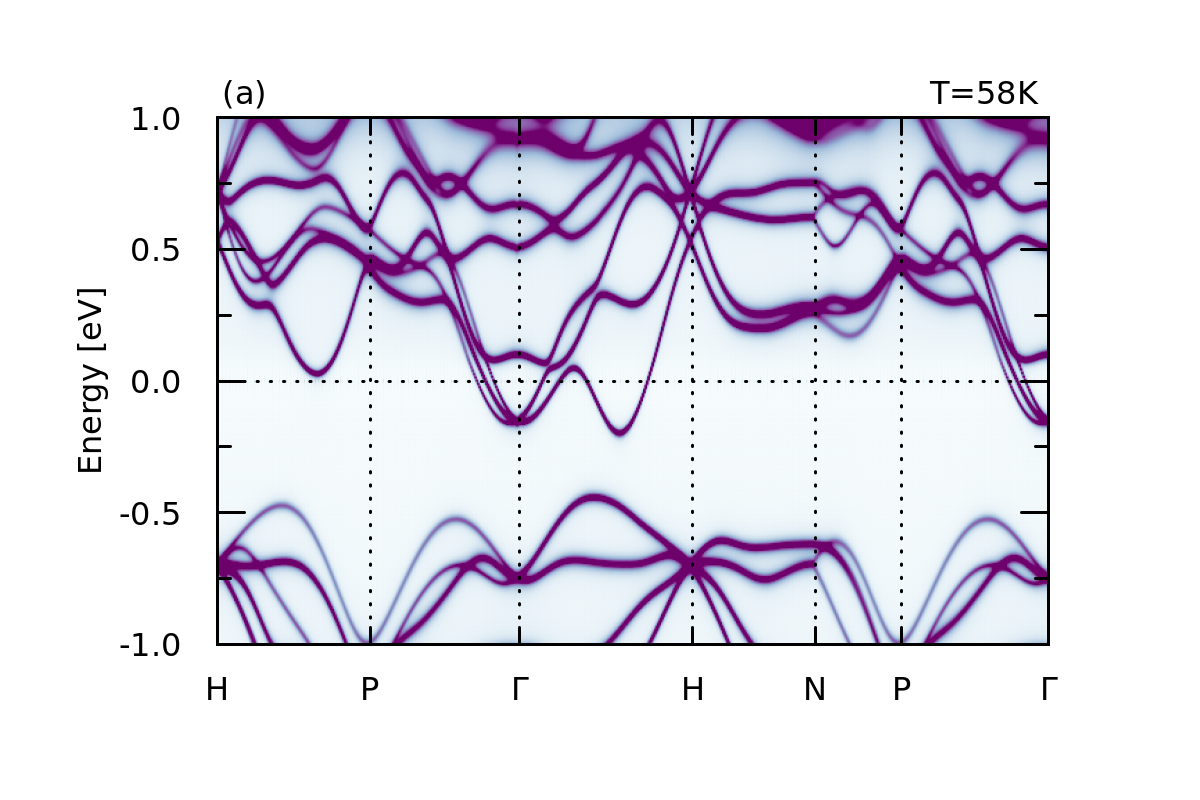}}
  \end{minipage}$\quad\quad$
   \begin{minipage}{0.49\textwidth}
  {\includegraphics[width=0.75\linewidth]{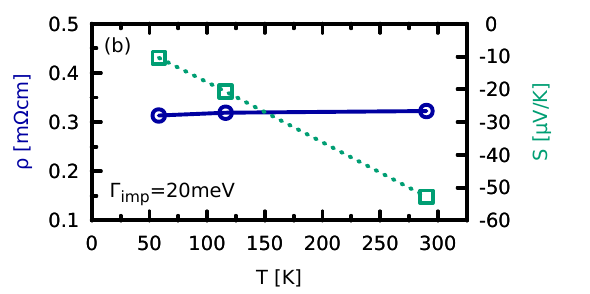}}
{\includegraphics[width=0.75\linewidth]{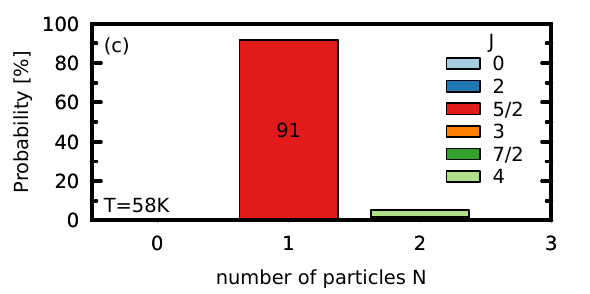}}
  \end{minipage}
  \caption{(a) Simulated many-body spectral function of Ce$_{3}$Bi$_{4}$Au$_{3}$ along a selected high-symmetry path in the Brillouin zone. 
  (b) resistivity $\rho$ (blue circles) and thermopower $S$ (green squares) simulated as a function of temperature, with an added impurity scattering rate $\Gamma_{\hbox{\tiny imp}}=20$~meV.
  (c) The valence histogram depicts the probability to find the $4f$ electrons of Ce 
  in an atomic-limit state with occupation $N$ and total angular momentum $J$. The data of (a) and (c) have been obtained at a temperature $T = 58$~K.}
  \label{DMFT}
  \end{figure*}

The $4f$ states of Ce hold the key to understanding the properties of Ce$_3$Bi$_4$Au$_3$. Depending on the compound, Ce$_3X_4T_3$, the physics can be very different:
The $4f^0$ reference material La$_3$Bi$_4$Au$_3$ is insulating within band-theory~\cite{Seibel16} whereas La$_3$Bi$_4$Pt$_3$ is metallic~\cite{NGCS,PhysRevResearch.4.L032004}. The extra $4f$ electron from cerium then leads---when neglecting electronic correlations---to a metallic band-structure in Ce$_3$Bi$_4$Au$_3$ (see Fig.~\ref{343DFT} in Appendix B), whereas Ce$_3$Bi$_4$Pt$_3$ exhibits a semiconducting (hybridization-)gap~\cite{doi:10.1143/JPSJ.62.2103,NGCS}. 

Appropriately treating many-body effects in the Kondo insulator Ce$_3$Bi$_4$Pt$_3$ results in a renormalization of the gap through an effective-mass enhancement of a factor of ten as well as in the emergence of Ce-$4f$ local moments above the lattice coherence temperature~\cite{NGCS,jmt_CBP_arxiv}.
In contrast, the many-body spectral function of Ce$_{3}$Bi$_{4}$Au$_{3}$, displayed in Figure~\ref{DMFT}a for $T=58$~K,
exhibits two electron pockets, one at the high-symmetry point $\Gamma$ and a slightly deeper one between $\Gamma$ and $H$, which corroborate the semi-metallic behavior observed in experiments. Interestingly, in the Kondo insulator Ce$_{3}$Bi$_{4}$Pt$_{3}$, the global minimum of the conduction band occurs at the $\Gamma$-point~\cite{jmt_CBP_arxiv,PhysRevLett.124.166403}. The difference between the spectra of Ce$_{3}$Bi$_{4}$Au$_{3}$ and Ce$_{3}$Bi$_{4}$Pt$_{3}$ thus goes beyond a mere shift of the chemical potential (to account for the additional electron per precious-metal atom). 
The energy of the two local minima (one at $\Gamma$ and one between $\Gamma-H$) in the dispersion has been shown to be strongly influenced by the position ($u$,$u$,$u$) of the Bi site in the unit-cell~\cite{LRT_Tstar}: a larger $u$ favors the band minimum to occur between $\Gamma$ and $H$. Consistent with this observation, experiments (simulations) for Ce$_{3}$Bi$_{4}$Au$_{3}$
find  $u=0.08642$ ($u=0.091$), which is larger than $u=0.084$~\cite{Severing91} ($u=0.088$) for Ce$_{3}$Bi$_{4}$Pt$_{3}$. In fact, already in the reference compound, La$_{3}$Bi$_{4}$Au$_{3}$, band-theory predicts the conduction-band minimum to be in between $\Gamma$ and $H$, see Fig.~\ref{343DFT}. The band-structure of La$_{3}$Bi$_{4}$Au$_{3}$ and the spectral function of Ce$_{3}$Bi$_{4}$Au$_{3}$
are comparable, with the crucial difference that in Ce$_{3}$Bi$_{4}$Au$_{3}$ the chemical potential has moved into the conduction band.

To understand this, we inspect the valence histogram in Fig.~\ref{DMFT}c, which displays the decomposition of the Ce-$4f$ electrons onto local states with well-defined occupation $N$ and total angular momentum $J$: Ce$_{3}$Bi$_{4}$Au$_{3}$ is dominated by states with $N=1$ and $J=5/2$. We find valence fluctuations, $\delta N=\sqrt{\left\langle \left( N-\langle N\rangle\right)^2\right\rangle}=0.29$, around the mean
Ce-$4f$ occupation $\langle N\rangle=1.07$ in Ce$_{3}$Bi$_{4}$Au$_{3}$ at $T=58$~K to be smaller than in the Kondo insulator Ce$_{3}$Bi$_{4}$Pt$_{3}$ (where $\delta N=0.36$ with $\langle N\rangle=1.02$~\cite{jmt_CBP_arxiv}).
Still, the presence of valence fluctuations indicates that the $4f$-electrons are not fully localized. Their participation in the bonding means that there is more itinerant charge available than in the $4f^0$ reference La$_3$Bi$_4$Au$_3$. As a consequence, the chemical potential has to move up into the conduction band.

The instantaneous local moment of Ce$_{3}$Bi$_{4}$Au$_{3}$,
$\mu_{inst}=
g_J\sqrt{\left\langle J^2\right\rangle}\mu_B/\hbar=g_J\sqrt{\left\langle j(j+1)\right\rangle}\mu_B\approx 2.65~\mu_B$
(using $g_{J=5/2}=0.857$),
is comparable to both the local moment of isolated Ce$^{3+}$ ions (see above) as well as to simulations of Ce$_{3}$Bi$_{4}$Pt$_{3}$ ($\mu_{inst}=2.64~\mu_B$~\cite{jmt_CBP_arxiv}).
However, the standard deviation, $\delta J=0.56$, of the total angular momentum is substantially smaller than $\delta J=0.81$ of Ce$_{3}$Bi$_{4}$Pt$_{3}$, again confirming the experimentally inferred stronger localization of the $f$-states in Ce$_{3}$Bi$_{4}$Au$_{3}$.
Finally, contrary to Ce$_{3}$Bi$_{4}$Pt$_{3}$ and Ce$_{3}$Bi$_{4}$Pd$_{3}$, the instantaneous local moment in Ce$_{3}$Bi$_{4}$Au$_{3}$ is hardly screened over time, resulting (when averaged over time) in a Curie-Weiss-like {\it local} magnetic susceptibility $\chi_{loc}(\omega=0)$, that is in excellent agreement with the experimental {\it uniform} susceptibility $\chi(q=0,\omega=0)$, see Fig.~\ref{ChivsT}.

Based on the (DMFT) electronic structure, we simulate transport properties, see Fig.~\ref{DMFT}b.
 The resistivity is only weakly temperature dependent---consistent with experiment and the semi-metallic spectrum.
Note that the calculations use an added temperature-independent scattering rate, $\Gamma_{\hbox{\tiny imp}}=20$~meV, to mimic sources of finite lifetimes other than electronic correlations, e.g., impurity scattering.
 The amplitude $\Gamma_{\hbox{\tiny imp}}$ is small compared to the scattering from electron-electron correlations: At the Fermi level, the Ce-$4f$ self-energy evaluates to $|\Im\Sigma(\omega=0)|=109$~meV ($184$~meV) at $T=58$~K ($T=290$~K) when averaged over the $J=5/2$ states.
 However, in Ce$_3$Bi$_4$Au$_3$, transport is mostly driven by the (non-$4f$) conduction electrons and the decay of their current is dominated by the impurity scattering (see Fig.~\ref{343transport}a in Appendix C).
 On an absolute scale, the simulated resistivity is too small by almost an order of magnitude compared to the experimental values.
This discrepancy could point to larger effects of disorder, e.g., structural defects or the impurity inclusions discussed above.
Indeed, the resistivity of Ce$_3$Sb$_4$Cu$_3$ was shown to vary by a factor of ten, depending on the sample preparation~\cite{WITAS2017256}. Similar effects could be responsible for the semimetallic behavior in the experimental resistivity of La$_3$Bi$_4$Au$_3$ in contrast to the DFT predicted insulating gap. Alternatively, off-stoichiometry could cause a shrinking of the electron pockets, leading to a larger resistivity.

Further insight comes from the Seebeck coefficient $S$.
As the ratio of two response functions~\cite{behnia}, $S$ is largely insensitive to the magnitude of scattering in metals%
\footnote{\jan{See, however, Ref.~\onlinecite{Sun_CoSb3,garmroudi2023high}.}}
(see Fig.~\ref{343transport}a). Its sign  provides information on the dominant type of charge carriers. Congruent with the presence of electron pockets in Fig.~\ref{DMFT}a, the simulated Seebeck coefficient in Fig.~\ref{DMFT}b is negative.
Future measurement of $S$ can thus serve as a validation of the simulated electronic structure.
If the thermopower of Ce$_3$Bi$_4$Au$_3$ turns out to be positive in experiments---as it is for Ce$_3$Sb$_4$Au$_3$~\cite{WITAS2017256} and Ce$_3$Sb$_4$Cu$_3$~\cite{10.1063/1.367018}---this might point towards extrinsic influences or electronic structure effects not captured here.

\section{Summary}

We have investigated the physical properties of cubic Ce$_{3}$Bi$_{4}$Au$_{3}$ by magnetic, thermodynamic, and electrical-transport measurements on single crystalline samples. Magnetization and heat-capacity data reveal that the 4$f$ moments in Ce$_{3}$Bi$_{4}$Au$_{3}$ are dominantly localized and order antiferromagnetically below $T_N=3.2$~K. The localized nature of the Ce moments and weak/negligible Kondo interactions are further evidenced by the slight enhancement of $T_N$ under the application of hydrostatic pressure. The electrical resistivity of Ce$_{3}$Bi$_{4}$Au$_{3}$ shows semimetallic behavior on cooling, which is corroborated by realistic many-body calculations that reveal the presence of small electron pockets below the Fermi level. In addition, the simulations quantitatively reproduce the local moment behavior in the paramagnetic state. Our combined experimental and theoretical investigations show that the additional electron per formula unit from Au atom compared to the Pt/Pd counterparts render Ce$_{3}$Bi$_{4}$Au$_{3}$ as a 4$f$-localized, semimetallic member of the highly tunable Ce$_3X_4T_3$ family of compounds.

\section*{Data Availability Statement}
Upon publication, the simulation data will be available at \href{https://doi.org/10.5281/zenodo.8298875}{doi.org/10.5281/zenodo.8298875}.

\section*{Acknowledgements}

We acknowledge constructive discussions with M.~M.~Bordelon. The conception of this work as well the crystal synthesis were supported by the U.S. Department of Energy, Office of Basic Energy Sciences ``Science of 100 T" program. J. Tomczak acknowledges support  by the Austrian Science Fund (FWF) through project BandITT Grant No.\ P 33571. Calculations were performed on the Vienna Scientific Cluster (VSC). M.~O.~Ajeesh acknowledges funding from the Laboratory Directed Research \& Development Program. Measurements under pressure were supported by the U.S. Department of Energy, Office of Basic Energy Sciences ``Quantum Fluctuations in Narrow Band Systems" program. Scanning electron microscope and energy dispersive x-ray measurements were performed at the Center for Integrated Nanotechnologies, an Office of Science User Facility operated for the U.S. Department of Energy Office of Science. The National High Magnetic Field Laboratory-Pulsed-Field Facility is funded by the National Science Foundation Cooperative Agreement Number DMR-1644779, the State of Florida and the U.S. Department of Energy. M.~K.~Chan acknowledges support from NSF IR/D program while serving at the National Science Foundation. Any opinion, findings, and conclusions or recommendations expressed in this material are those of the author(s) and do not necessarily reflect the views of the National Science Foundation.

\section*{appendix}
\subsection{Magneto-transport: Effect of impurity inclusions}
\begin{figure}[tb]
\centering
\includegraphics[width=1\linewidth]{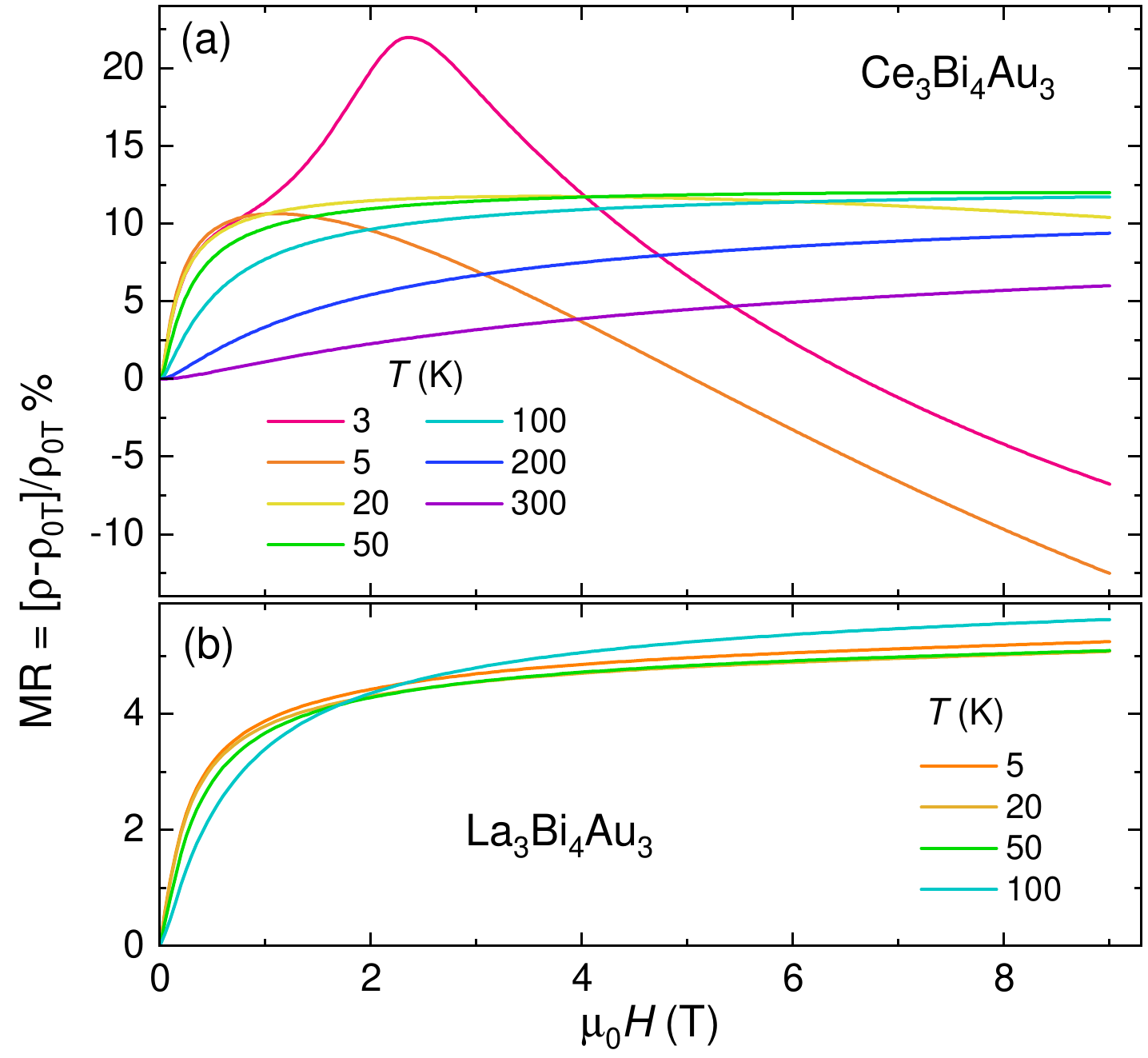}
\caption{ Transverse Magnetoresistance MR$=[\rho-\rho_{\rm 0}]/\rho_{\rm0}$ as a function of magnetic field measured at different temperatures for (a) Ce$_{3}$Bi$_{4}$Au$_{3}$ and (b) La$_{3}$Bi$_{4}$Au$_{3}$.}
\label{MRvsH}
\end{figure}

The transverse magnetoresistance MR$=[\rho-\rho_{\rm 0T}]/\rho_{\rm 0T}$ as a function of magnetic field measured at different temperatures for Ce$_{3}$Bi$_{4}$Au$_{3}$ and La$_{3}$Bi$_{4}$Au$_{3}$ is shown in Fig.~\ref{MRvsH}. MR($H$) of both compounds show very similar behavior apart from the contributions from the magnetic transitions in Ce$_{3}$Bi$_{4}$Au$_{3}$. The sharp increase in MR for fields up to 1~T is clearly seen in both compounds and is nearly temperature independent up to 50~K. In addition the increase in MR also shows strong sample dependence for both the compounds. All these results point to an extrinsic origin for the behavior, possibly due to the presence of Bi or Au-Bi inclusions in the samples. The resistivity data on all samples show superconducting transitions at low temperatures coming from tiny inclusions of Au-Bi binaries. These inclusions may be present in the form of irregular filaments that would give rise to current inhomogeneities. In addition, the inclusions could range from elemental bismuth to Bi-Au binary compounds of varying stoichiometry. The normal state of such binaries often have semimetallic behavior with strong magnetoresistance properties. In the case of elemental Bi, strongly anisotropic magneto-transport behavior is observed along with a semimetal to insulator-like behavior change under magnetic field applied along the trigonal axis of Bi single crystal~\cite{Du05}. Effects can be further complicated as the inclusions can have geometries such as thin-layer or filament shape which could lead to different properties compared to bulk samples~\cite{Chu92, Chu88, Lu96}. Notably, a very similar behavior in MR for fields up to 1~T is reported in flux grown samples of (Ce,La)$_{3}$Bi$_{4}$Pt$_{3}$~\cite{Hundley93}. Even though the authors used a multi-band model to explain the MR data, the striking similarity to the MR of (Ce,La)$_{3}$Bi$_{4}$Au$_{3}$ with a significantly different electronic band structure (induced by the additional electron per formula unit from Au compared to Pt) is rather suspicious. A more plausible explanation is the effect of semimetallic inclusions common to all these materials.

\subsection{Density functional theory}
\begin{figure}[tb]
\centering
\includegraphics[angle=0,width=1\linewidth]{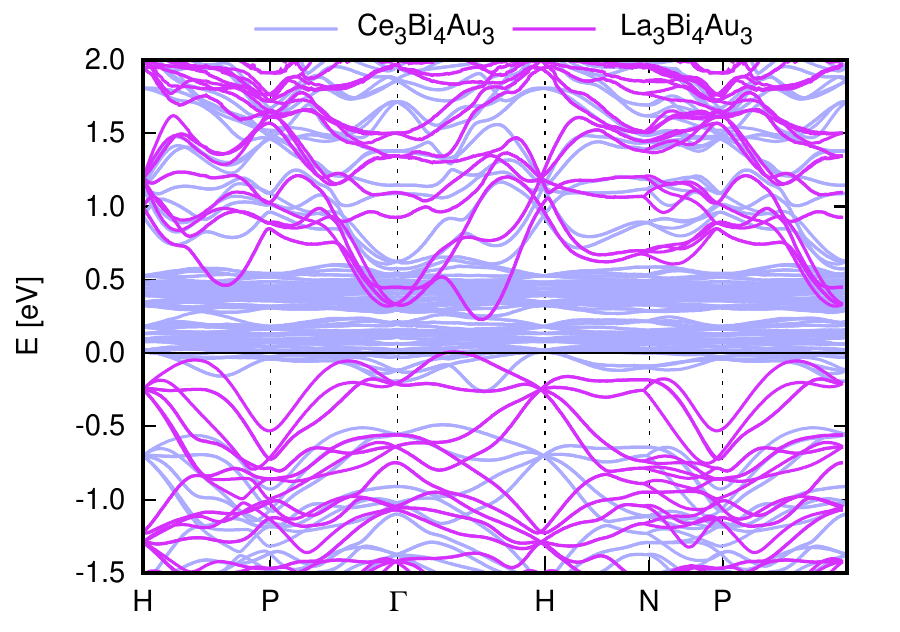}
\caption{Band-structures of Ce$_{3}$Bi$_{4}$Au$_{3}$ (light blue) and La$_{3}$Bi$_{4}$Au$_{3}$ (magenta; see also Ref.~\cite{Seibel16}) within DFT: The system is insulating (metallic) in the absence (presence) of the $4f$-electron---in direct opposition to the related pair of materials RE$_3$Bi$_4$Pt$_3$ (RE=Ce,La)\cite{NGCS}.}
\label{343DFT}
\end{figure}
In Fig.~\ref{343DFT} we compare the band-structures of 
 La$_{3}$Bi$_{4}$Au$_{3}$ (insulating; magenta) and Ce$_{3}$Bi$_{4}$Au$_{3}$ (metallic; blueish).
 The spectral function of Ce$_{3}$Bi$_{4}$Au$_{3}$ in Fig.~\ref{DMFT}a incorporates many-body effects within DMFT and is shown for the same path in the Brillouin zone.
As the Ce-$4f$ states are mostly localized, the many-body spectrum of Ce$_{3}$Bi$_{4}$Au$_{3}$ is reminiscent of the La$_{3}$Bi$_{4}$Au$_{3}$ band-structure---albeit with a chemical potential that is shifted into the conduction band owing to finite hybridization with the $4f$-states.
 
\subsection{Transport simulations}
\begin{figure}[tb]
\centering
\includegraphics[angle=0,width=1\linewidth]{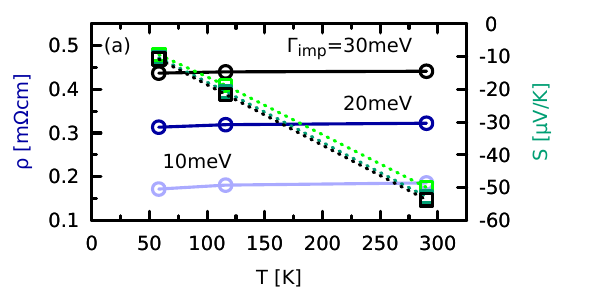}
\includegraphics[angle=0,width=1\linewidth]{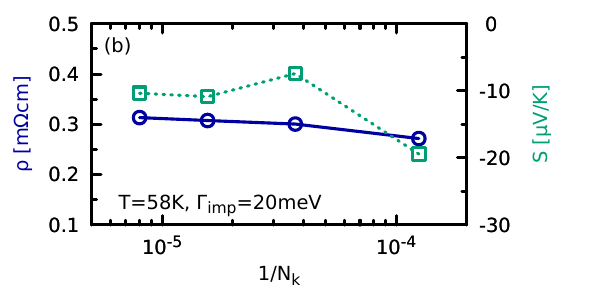}
\caption{Simulated resistivity and Seebeck coefficient: (a) dependence on the added impurity scattering rate (light to dark: $\Gamma_{\hbox{\tiny imp}}=10$, 20, 30~meV); blue: resistivity $\rho$; green: Seebeck coefficient $S$. (b) convergence of $\rho$ and $S$ with the Brillouin zone discretization ($N_k$: number of reducible $\mathbf{k}$ points). The calculations in Fig.~\ref{DMFT}b use the finest mesh ($N_k=50^3$).}
\label{343transport}
\end{figure}

As alluded to in the main text, the Seebeck coefficient $S$ of metals is virtually insensitive to (reasonable) changes in the scattering rate $\Gamma$, see Fig.~\ref{343transport}a. The resistivity $\rho$ of metals is, instead, basically proportional to $\Gamma$. In the case of Ce$_{3}$Bi$_{4}$Au$_{3}$, it is $\Gamma_{\hbox{\tiny imp}}$ that dominates the response of the conduction electrons. Transport simulations of (semi-)metals are delicate. In Fig.~\ref{343transport}b, we therefore assess the convergence of the response with respect to the Brillouin zone discretization. At the lowest temperature considered here, $T=58$~K, and for $\Gamma_{\hbox{\tiny imp}}=20$~meV, both $\rho$ and $S$ appear reasonably converged for a momentum grid with $50^3$ reducible points.

\pagebreak

\bibliography{Ce3Bi4Au3}

\end{document}